\documentclass{amsart}

\newcommand{\pictures}{}

\usepackage[
	pdftitle={Hard edge},
	pdfauthor={Leonid  O. Chekhov},
	ocgcolorlinks,
	linkcolor=linkblue,
	citecolor=linkred,
	urlcolor=linkred]
{hyperref}

\usepackage{amsfonts}
\usepackage{amssymb}
\usepackage{bbm}
\usepackage{bm}
\usepackage{booktabs}
\usepackage[hang,flushmargin]{footmisc}
\usepackage[left=25mm,right=25mm,top=25mm,bottom=25mm]{geometry}
\usepackage{graphicx}
\usepackage{mathpazo}
\usepackage{microtype}
\usepackage{multicol}
\usepackage{tikz}

\def\appendix#1{
\addtocounter{section}{1} \setcounter{equation}{0}
\renewcommand{\thesection}{\Alph{section}}
\section*{Appendix \thesection\protect\indent\quad
#1}
}
\def\beq{\begin{equation}}
\def\eeq{\end{equation}}

\usepackage{color}

\ifdefined\pictures
\usepackage[pdf]{pstricks}
\usepackage{pst-node}
\usepackage{pst-plot}
\fi

\usepackage{mathtools}

     \usetikzlibrary{arrows,shapes}
\usepackage{xcolor}
     \definecolor{linkred}{rgb}{0.6,0,0}
     \definecolor{linkblue}{rgb}{0,0,0.6}

\theoremstyle{plain}
     \newtheorem{theorem}{Theorem}
     \newtheorem{lemma}{Lemma}
     \newtheorem{proposition}{Proposition}[section]
     
     \newtheorem{corollary}[proposition]{Corollary}

\theoremstyle{definition}
     \newtheorem{example}[proposition]{Example}
     \newtheorem{definition}[proposition]{Definition}
     \newtheorem{remark}[proposition]{Remark}

\newcommand{\res}{\mathop{\mathrm{Res}}}
\newcommand{\bb}{\boldsymbol{b}}
\newcommand{\pp}{\boldsymbol{p}}
\newcommand{\bc}{\mathbb{C}}
\newcommand{\bn}{\mathbb{N}}
\newcommand{\bp}{\mathbb{P}}
\newcommand{\bz}{\mathbb{Z}}

\newcommand{\ch}{\mathcal{H}}
\newcommand{\cl}{\mathcal{L}}
\newcommand{\modm}{\mathcal{M}}

\newcommand{\cp}{\mathcal{P}}
\newcommand{\ct}{\mathcal{T}}

\newcommand {\dd}{\mathrm{d}}

\newcommand {\zz}{\bm{z}}

\newcommand{\un}{1\!\!1}
\newcommand{\Res}{\mathop{\,\rm Res\,}}

\newcommand*{\Cdot}{\raisebox{-0.5ex}{\scalebox{1.8}{$\cdot$}}} 

\newcommand{\tcr}{\textcolor{red}}
\newcommand{\tcb}{\textcolor{blue}}

\def\tr{{\mathrm{tr\,}}}

\numberwithin{equation}{section}

\def\be{\begin{equation}}
\def\ee{\end{equation}}

\long\def\rem#1{}

\begin{document}

\title{Virtual Euler characteristics via topological recursion}
\author{Leonid O. Chekhov}
\address{Steklov Mathematical Institute and Michigan State University, East Lansing, USA}
\email{\href{mailto:chekhov@mi-ras.ru}{chekhov@mi-ras.ru}}
\thanks{The author thanks Paul Norbury who participated at early stages of this project for numerous valuable remarks.}
\subjclass[2010]{14N10; 05A15; 32G15}
\date{\today}

\begin{abstract}
We use Seiberg--Witten-like relations in the topological recursion framework to obtain virtual Euler characteristics for uni- and multicellular maps for ensembles of classic  orthogonal polynomials and for ensembles related to nonorientable surfaces. We also discuss Harer--Zagier-type recursion relations for 1-point correlation function for the Legendre ensemble. 
\end{abstract}

\maketitle

\setlength{\parskip}{0pt}
\setlength{\parskip}{6pt}

\section{Introduction}  \label{s:intro}

In this note, we probe virtual Euler characteristics for uni- and multicellular maps for classical ensembles: Hermite, Laguerre, and Legendre. We claim that, under the proper definition of the virtual Euler characteristics, we can effectively evaluate it using the Seiberg--Witten-like relations.

Whereas a remarkable progress was achieved in the last decade in solving enumerations problems in a number of models using a 
\emph{topological recursion} technique developed in \cite{CEyHer}, \cite{EOrInv} and independently in \cite{AMMBGW}, \cite{AMMIns}, recursion relations underlying this technique are nonlinear, which makes it hard to go beyond several first iterations of genus expansion.

On the other hand, in a handful of cases, we have linear recursion relations enjoyed by one-point resolvents $W_1^{(g)}(x)$ of the corresponding models. The first example of such relation was found by Harer and Zagier \cite{HZ86} for the Hermitian matrix model (Gaussian Unitary Ensemble, or GUT) and the authors used their recursion to find \emph{virtual Euler characteristic} of moduli space $\overline{\mathcal M}_{g,s}$ stratified by Deligne and Mumford. Almost simultaneously, Penner introduced his matrix model \cite{Penner} evaluating the same characteristics directly. Since then, the list of models admitting linear differential equations satisfied by the one-point resolvents enlarged to incorporate the classic Laguerre ensemble \cite{Chap11}, \cite{CFF}, generalized Laguerre ensembles \cite{Ch16}, classic Legendre ensemble \cite{GKR05}, and Hermitian $\beta$-model ensembles for $\beta=1,4$ (GOE and GSE) \cite{Ledoux}, \cite{WitteForrester}.

Note however that inside the topological recursion framework we have exact linear relations commonly called Seiberg--Witten relations that hold in all orders of the genus expansion \cite{CMMV2}. Whereas using these relations for producing systems of linear differential equations on $W_1^{(g)}(x)$ does not seem feasible, the aim of this paper is to show how we can find the virtual Euler characteristics of various models using the linear Seiberg--Witten relations. 

Our main goal is to compare results that we obtain for the virtual Euler characteristics $\varkappa_{g,s}$ in various matrix models using three methods:
\begin{itemize}
\item The Sieberg--Witten relations, $\varkappa_{g,s}=\frac{1}{s!}\frac{\partial^s}{\partial t_0^3}F_g^{\text{Free}}(t_0)$, where we set the number of eigenvalues to be equal $t_0N$.
\item The direct evaluation of $\varkappa_{g,s}$ using Penner-like matrix models with logarithmic potentials,   $\varkappa_{g,s}=\sum_{\Gamma_{g,s}}\frac{(-1)^{\#\,\text{edges}}}{|\text{Aut}\,\Gamma_{g,s}|}$, where the sum ranges fat graphs of genus $g$ with $s$ faces with vertices of arbitrary orders. 
\item The Givental-like decomposition: We express connected correlation functions $W^{(g)}_s(T^{\pm}_k(\lambda))$ in terms of the KdV hierarchy times $T^{\pm}_k$ related to local models situated at the two (in $\mathbb P^1$ case) branching points; the virtual Euler characteristics is then merely
$\varkappa_{g,s}=W^{(g)}_s\bigm|_{T^{\pm}_k=\delta_{k,0}}$.
\end{itemize}
Despite the fact that an explicit combinatorial proof of equivalence of all three methods exists, to the best of our knowledge only for the GUT and is presented in this note, in many, but not all, situations where we can evaluate $\varkappa_{g,s}$ using different methods, the results coincide.

Notably, there is a fourth way to calculate $\varkappa_{g,s}$, at least, for some models: using algebraic-geometric tools related to integrals of Chern, Hodge, and other classes over the corresponding moduli spaces; using these tools, Norbury  calculated \cite{Nor18} virtual Euler characteristics $\varkappa_{g,1}$ in the Legendre model case, and Giacchetto, Lewanski, and Norbury reproduced \cite{GLN21} the classic result of Harer and Zagier for standard GUT $\varkappa_{g,s}$.

Technically, the first method is the simplest one; in order to apply it we only need to know the exact analytic in $t_0$ expression for the partition function of a free model; the second method is harder, but still manageable, whereas the third seems to be superhard technically, and we were able to compare answers only in several lowest orders of genus expansion for the Legendre model.

In this paper, we calculate $\varkappa_{g,s}$ for four models: the GUT, the Legendre model, the model of symmetric real matrices (GOE, or $\beta=1$-model), and  the model of rectangular complex matrices $B\in \mathbb C^{(1+\gamma)N\times N}$ (the Marchenko--Pastur ensemble \cite{MP}) using the first two methods whenever possible.  We collect our results into a table  indicating in the last column whether the results obtained using different techniques coincide or not.
\begin{table}[htp]
\begin{center}
\begin{tabular}{|c|c|c|c|c|}
\hline
$F_g^{\text{Free}}(t_0)$ & Log-potential & $F_g(T^\pm_k)(\lambda)$ & Alg-geom. & Coincide? \\
\hline
\hline
GUT & {Gener. Laguerre (Penner)} & $e^{\mathcal A}Z_{KW}\otimes Z_{KW}$ & \cite{GLN21} & Yes \\
\hline
Legenrde Ensemble & ? & $e^{\mathcal A}Z_{BGW}\otimes Z_{BGW}$ & \cite{Nor18} & Yes \\
\hline
{GOE ($\beta=1$ Hermitian)} & {GOE ($\beta=1$ Laguerre)} & {presumably DNE} & ? & Yes \\
\hline
{Marchenko--Pastur ensemble} & Gener. Legendre & $e^{\mathcal A}Z_{BGW/KW}\otimes Z_{KW}$ & ? & No \\
\hline
\end{tabular}
\end{center}
\label{default}
\end{table}%

We begin with a brief accounting for the topological recursion in its original formulation: it produces genus-filtrated invariants of a Riemann surface $\Sigma$ equipped with two meromorphic functions $x, y: \Sigma\to \mathbb{C}$ and a bidifferential $B(p_1,p_2)$ for $p_1, p_2 \in \Sigma$.  The zeros of $dx$ are assumed to be simple.  The data $(\Sigma,B,x,y)$ is commonly referred to as a {\em spectral curve}.   For integers $g \geq 0$ and $n \geq 1$, the ``correlation function'' $W^{(g)}_{n}$ is a $n$-tuple totally symmetric differential on $\Sigma$, or a tensor product of meromorphic differentials on $\Sigma^n$.  It is defined recursively via the initial conditions
$$
W_{1}^{(0)}(p)=-y(p)dx(p),\quad W_{2}^{(0)}(p_1,p_2)=B(p_1,p_2)
$$
which are used to define the kernel $K$, which is a $(1,-1)$-differential, in a neighbourhood of $p_2=\alpha$ for $dx(\alpha)=0$
$$ 
K(p_1,p_2)=\frac{1}{2}\frac{\int_{\hat{p}_2}^{p_2}B(p,p_1)}{(y(p_2)-y(\hat{p}_2))dx(p_2)}.
$$
The point $\hat{p}\in \Sigma$ is the unique point $\hat{p}\neq p$ close to $\alpha$ such that $x(\hat{p})=x(p)$, which is well-defined since each zero $\alpha$ of $\dd x$ is assumed to be simple.  For $S=\{2,\dots,n\}$ define
\begin{equation}  \label{TRrec}
W_{n}^{(g)}(p_1,\pp_{S})=\sum_{\alpha}\res_{p=\alpha}K(p_1,p) \bigg[W_{n+1}^{(g-1)}(p,\hat{p},\pp_{S})+ \mathop{\sum_{g_1+g_2=g}}_{I\sqcup J=S}^\circ W_{|I|+1}^{(g_1)}(p,\pp_I) \, W_{|J|+1}^{(g_2)}(\hat{p},\pp_J) \bigg]
\end{equation}
where the outer summation is over the zeros $\alpha$ of $dx$ and the $\circ$ over the inner summation means that we exclude terms that involve $\omega_1^0$.  A zero of $dx$ is {\em regular} if $y$ is analytic there.   A spectral curve is regular if $y$ is analytic at all zeros of $dx$.  {\em Irregular} spectral curves correspond to cases where $y$ has simple poles at some zeros of $dx$. If $y$ has higher order poles, these points drop  out of recursion procedure as residues at these points in the recursion (\ref{TRrec}) always vanish. 

Correlation functions $W_{n}^{(g)}$ enjoy {\bf exact}, that is, valid in each order of the topological expansion separately, Seiberg--Witten equations (see \cite{Kri92},  \cite{CMMV1},  \cite{CMMV2} and references therein). The set of flat variables comprises $t_k$---the times of the potential, $a_\alpha:=\oint_{A_\alpha}ydx$---the occupation numbers, and $t_0:=\hbox{res}_\infty ydx$---the normalized number of eigenvalues, which is perhaps the most "overseen" among all flat coordinates. The Seiberg--Witten equations read:
\be\label{SW-A}
\frac{\partial F_g}{\partial a_\alpha}=\oint_{B_\alpha}W_1^{(g)},\qquad \frac{\partial W_k^{(g)}}{\partial a_\alpha}=\oint_{B_\alpha}W_{k+1}^{(g)}
\ee
\be\label{SW-t0}
\frac{\partial F_g}{\partial t_0}=\int_{\infty}^{\overline\infty}W_1^{(g)},\qquad \frac{\partial W_k^{(g)}}{\partial t_0}=\int_{\infty}^{\overline\infty}W_{k+1}^{(g)},
\ee
where $\overline\infty$ is the copy of the infinite point on the other (nonphysical) sheet of the spectral curve. In particular, we have
\be\label{SW-t0-mult}
\frac{\partial^k F_g}{\partial t_0^k}=\underbrace{\int_{\infty}^{\overline\infty}\cdots \int_{\infty}^{\overline\infty}}_kW_k^{(g)},\quad 2g-2+k>0.
\ee
In this short note, we show that this relation produces virtual Euler characteristics of moduli spaces of all classic matrix model ensembles, and we postulate that it produces virtual Euler characteristics also in cases where a direct geometrical description is lacking, which is the case, e.g., for moduli spaces of nonoriented surfaces. 

\section{Virtual Euler characteristic for the Gaussian model}

The virtual Euler characteristics $\varkappa_{g,s}$ of the moduli space $\overline{\mathcal M}_{g,s}$ compactified by Deligne and Mumford is an alternative-sign sum over strata of the corresponding cell decomposition with symmetries of the cells taken into account by the reciprocal volumes of automorphism groups $(-1)^E/\#\hbox{Aut\,}\Gamma$. For Poincar\'e uniformized curves, strata that give nonzero contribution are in one-to-one correspondence with fat graphs of genus $g$, with $s\ge 1$ faces, and with $E$ edges, $2g-1+s\le E\le 6g-6+3s$, where $E$ defines the dimension of the corresponding cell. Other strata that correspond to reduced curves have automorphism groups of infinite volumes and therefore drop out of the expression for the virtual Euler characteristics.

So in the case of Gaussian model, the virtual Euler characteristics $\varkappa_{g,s}$ for the moduli space of $s$-component ($s$ marked points) genus $g$ surfaces is given by the following finite sum over proper connected fat graphs, i.e., those with vertices of valence three and higher:
\be\label{kgs}
\varkappa_{g,s}
=\sum_{{\mathrm{all\ genus\ } g \mathrm{\ proper}}\atop \mathrm{fat\ graphs\ }\Gamma\mathrm{\ with\ }s\mathrm{\ faces}}
\frac{1}{\#\text{Aut\,} (\Gamma)} (-1)^{\#\text{vertices}}.
\ee
These are expansion terms for the free energy of the logarithmic matrix model introduced by R.Penner \cite{Penner}. 

Let $M$ be an $N\times N$ Hermitian matrix and $DM=\prod_{i}dM_{i,i}\prod_{i<j}d \Re M_{i,j} d\Im M_{i,j}$ be the standard Haar measure of integration. Then the correlation function 
$$
\langle M_{i,j} M_{k,l}\rangle :=\frac{1}{Z_0}\int M_{i,j} M_{k,l} e^{-\frac{N}{2}\tr M^2}DM=\delta_{j,k}\delta_{i,l}\sim
\frac {1}{N}\raisebox{-12pt}{\mbox{
\begin{pspicture}(-0.7,-0.5)(0.7,0.5)
\pcline[linewidth=12pt,linestyle=solid](-0.5,0)(0.5,0)
\pcline[linewidth=10pt,linecolor=white,linestyle=solid](-0.52,0)(0.52,0)
\rput(-0.55,0.3){\makebox(0,0)[cr]{$i$}}
\rput(-0.55,-0.3){\makebox(0,0)[cr]{$j$}}
\rput(0.55,0.3){\makebox(0,0)[cl]{$l$}}
\rput(0.55,-0.3){\makebox(0,0)[cl]{$k$}}
%
\end{pspicture} 
}}
$$
can be depicted as an edge of a fat graph with indices running along the sides; the terms $\frac 1n \tr M^n$ then can be depicted as fat-graph vertices of order $n$, e.g., for $n=3$, we have
$$
-\frac N3 \, \tr M^3=-\frac{N}{3}\sum_{i,j,k} M_{i,j} M_{j,k}M_{k,i}\sim -N \raisebox{-14pt}{\mbox{
\begin{pspicture}(-0.9,-0.7)(0.9,0.7)
\psarc[linewidth=1pt,linecolor=black](1,0){0.7}{140}{220}
\rput{120}(0,0){\psarc[linewidth=1pt,linecolor=black](1,0){0.7}{140}{220}}
\rput{240}(0,0){\psarc[linewidth=1pt,linecolor=black](1,0){0.7}{140}{220}}
\rput(-0.7,0.3){\makebox(0,0)[cr]{$i$}}
\rput(0,-0.5){\makebox(0,0)[ct]{$j$}}
\rput(0,0.5){\makebox(0,0)[cb]{$i$}}
\rput(-0.7,-0.3){\makebox(0,0)[cr]{$j$}}
\rput(0.5,0.4){\makebox(0,0)[cl]{$k$}}
\rput(0.5,-0.4){\makebox(0,0)[cl]{$k$}}
%
\end{pspicture} 
}}
$$
with the combinatorial factor $1/n$ taking care on the cyclic symmetry of the vertex. The matrix-model potential with the added Gaussian term then becomes $N\sum_{n=2}^\infty \frac1n \tr M^n=\tr \bigl(\log(1-M) +M\bigr)$ and upon changing $M\to-H$ we obtain the Penner matrix integral (\ref{Penner}) below. Note that index lines constitute faces of the fat graph, so any connected fat graph $\Gamma$ contribute the total factor  $\frac{1}{\#\text{Aut\,} (\Gamma)} (-1)^{\#\text{vertices}}N^{\#\text{vertices}+\#\text{faces}-\#\text{edges}}$ with the power of $N$ being an Euler characteristics of $\Gamma$. 


\subsection{Representing Gaussian model in canonical times}
We first recall the Catalan number calculus related to summing up rainbow and ladder contributions (see \cite{ACNP}). 

Consider a sum of connected diagrams with $s$ backbones (a backbone is a term $\tr H^k$ weighted by $1/k$) each carrying the corresponding variable 
\be
x_i:=e^{\lambda_i}+e^{-\lambda_i},\quad i=1,\dots,s.
\ee

Starting with the connected correlation function
\be
\label{det-rep}
\left\langle\prod_{i=1}^s\tr\log(x_i-H)\right\rangle_g^{\mathrm{conn}},
\ee
averaged over the Gaussian ensemble with the measure $e^{-\frac12 \tr H^2}$ with only connected diagrams of genus-$g$ topological type taken into account, we obtain multiloop means (symmetric $s$-tuple differentials) $W_s^{(g)}(\lambda_\bullet)$ by differentiation:
\begin{align}
W_s^{(g)}(\lambda_\bullet)&=\prod_{i=1}^s \frac{\partial}{\partial x_i} \left\langle\prod_{i=1}^s\tr\log(x_i-H)\right\rangle_g^{\mathrm{conn}}dx_1\cdots dx_s\nonumber\\
&=\prod_{i=1}^s \frac{\partial}{\partial \lambda_i} \left\langle\prod_{i=1}^s\tr\log(e^{\lambda_i}+e^{-\lambda_i}-H)\right\rangle_g^{\mathrm{conn}}d\lambda_1\cdots d\lambda_s
\label{loop-rep}
\end{align}

We first make partial summations in (\ref{det-rep}) of ``rainbow'' diagrams, which is merely the Catalan number counting and effectively reduces to replacing the original propagator $x_i^{-2}=1/(e^{\lambda_i}+e^{-\lambda_i})^2$ by $1+e^{-2\lambda_i}$ for any segment of a backbone that  carries the variable $x_i$ and becomes disjoint from the rest of the diagram if we cut its two bounding edges; we indicate this new propagators by double edges.

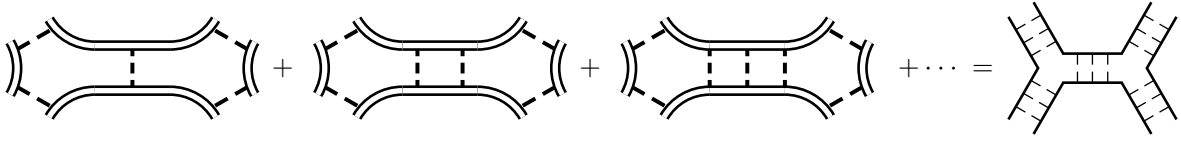
\begin{figure}[h]
{\psset{unit=1}
\begin{pspicture}(-2,-1)(2,1)
\newcommand{\PATTERN}{%
\rput(-0.6,0.25){\pcline[linewidth=1.5pt,linestyle=dashed](0,0.25)(-0.433,0)}
\rput(-0.6,-0.25){\pcline[linewidth=1.5pt,linestyle=dashed](0,-0.25)(-0.433,0)}
\psarc[linewidth=4pt,linecolor=black](0,1){0.7}{210}{270}
\psarc[linewidth=2pt,linecolor=white](0,1){0.7}{210}{270}
\psarc[linewidth=4pt,linecolor=black](0,-1){0.7}{90}{150}
\psarc[linewidth=2pt,linecolor=white](0,-1){0.7}{90}{150}
\psarc[linewidth=4pt,linecolor=black](-1.732,0){0.7}{-30}{30}
\psarc[linewidth=2pt,linecolor=white](-1.732,0){0.7}{-30}{30}
}
\pcline[linewidth=1.5pt,linestyle=dashed](0,-0.3)(0,0.3)
\pcline[linewidth=4pt,linestyle=solid](-0.5,0.3)(0.5,0.3)
\pcline[linewidth=2pt,linecolor=white,linestyle=solid](-0.5,0.3)(0.5,0.3)
\pcline[linewidth=4pt,linestyle=solid](-0.5,-0.3)(0.5,-0.3)
\pcline[linewidth=2pt,linecolor=white,linestyle=solid](-0.5,-0.3)(0.5,-0.3)
\rput(-0.5,0){\PATTERN}
\rput{180}(0.5,0){\PATTERN}
\rput(2,0){\makebox(0,0)[cc]{$+$}}
%
\end{pspicture} 
\begin{pspicture}(-2,-1)(2,1)
\newcommand{\PATTERN}{%
\rput(-0.6,0.25){\pcline[linewidth=1.5pt,linestyle=dashed](0,0.25)(-0.433,0)}
\rput(-0.6,-0.25){\pcline[linewidth=1.5pt,linestyle=dashed](0,-0.25)(-0.433,0)}
\psarc[linewidth=4pt,linecolor=black](0,1){0.7}{210}{270}
\psarc[linewidth=2pt,linecolor=white](0,1){0.7}{210}{270}
\psarc[linewidth=4pt,linecolor=black](0,-1){0.7}{90}{150}
\psarc[linewidth=2pt,linecolor=white](0,-1){0.7}{90}{150}
\psarc[linewidth=4pt,linecolor=black](-1.732,0){0.7}{-30}{30}
\psarc[linewidth=2pt,linecolor=white](-1.732,0){0.7}{-30}{30}
}
\pcline[linewidth=1.5pt,linestyle=dashed](-0.3,-0.3)(-0.3,0.3)
\pcline[linewidth=1.5pt,linestyle=dashed](0.3,-0.3)(0.3,0.3)
\pcline[linewidth=4pt,linestyle=solid](-0.5,0.3)(0.5,0.3)
\pcline[linewidth=2pt,linecolor=white,linestyle=solid](-0.5,0.3)(0.5,0.3)
\pcline[linewidth=4pt,linestyle=solid](-0.5,-0.3)(0.5,-0.3)
\pcline[linewidth=2pt,linecolor=white,linestyle=solid](-0.5,-0.3)(0.5,-0.3)
\rput(-0.5,0){\PATTERN}
\rput{180}(0.5,0){\PATTERN}
\rput(2,0){\makebox(0,0)[cc]{$+$}}
\end{pspicture} 
\begin{pspicture}(-2,-1)(3,1)
\newcommand{\PATTERN}{%
\rput(-0.6,0.25){\pcline[linewidth=1.5pt,linestyle=dashed](0,0.25)(-0.433,0)}
\rput(-0.6,-0.25){\pcline[linewidth=1.5pt,linestyle=dashed](0,-0.25)(-0.433,0)}
\psarc[linewidth=4pt,linecolor=black](0,1){0.7}{210}{270}
\psarc[linewidth=2pt,linecolor=white](0,1){0.7}{210}{270}
\psarc[linewidth=4pt,linecolor=black](0,-1){0.7}{90}{150}
\psarc[linewidth=2pt,linecolor=white](0,-1){0.7}{90}{150}
\psarc[linewidth=4pt,linecolor=black](-1.732,0){0.7}{-30}{30}
\psarc[linewidth=2pt,linecolor=white](-1.732,0){0.7}{-30}{30}
}
\pcline[linewidth=1.5pt,linestyle=dashed](-0.5,-0.3)(-0.5,0.3)
\pcline[linewidth=1.5pt,linestyle=dashed](0,-0.3)(0,0.3)
\pcline[linewidth=1.5pt,linestyle=dashed](0.5,-0.3)(0.5,0.3)
\pcline[linewidth=4pt,linestyle=solid](-0.5,0.3)(0.5,0.3)
\pcline[linewidth=2pt,linecolor=white,linestyle=solid](-0.5,0.3)(0.5,0.3)
\pcline[linewidth=4pt,linestyle=solid](-0.5,-0.3)(0.5,-0.3)
\pcline[linewidth=2pt,linecolor=white,linestyle=solid](-0.5,-0.3)(0.5,-0.3)
\rput(-0.5,0){\PATTERN}
\rput{180}(0.5,0){\PATTERN}
\rput(2,0){\makebox(0,0)[lc]{$+\cdots\ =$}}
\end{pspicture} 
\begin{pspicture}(-1.5,-1)(1.5,1)
\newcommand{\PATTERN}{%
\pcline[linewidth=12pt,linestyle=solid](0.1,0)(0.9,0)
\pcline[linewidth=10pt,linecolor=white,linestyle=solid](0.05,0)(0.95,0)
\multiput(0.3,0)(0.2,0){3}{\pcline[linewidth=0.5pt,linestyle=dashed](0,0.2)(0,-0.2)}
}
\rput{120}(-0.5,0){\PATTERN}
\rput{240}(-0.5,0){\PATTERN}
\rput{60}(0.5,0){\PATTERN}
\rput{-60}(0.5,0){\PATTERN}
\rput(-0.5,0){\PATTERN}
\end{pspicture} 
}
\caption{\small Doing a sum over ladder diagrams. New propagators are depicted as tiny ladders.}
\label{fi:cycles}
\end{figure}

We now sum up ladder diagrams. Here, we joint two cycles composed out of double edges and carrying in general different but possibly coinciding variables $\lambda_i$ and $\lambda_j$ by $\langle HH\rangle$ propagators (with the weights $x_i^{-1}x_j^{-1}$) interlaced with the double edges of the two cycles (see Fig.~\ref{fi:cycles}). We then collapse segments of two cycles into a new propagator carrying two numbers $\lambda_i$ and $\lambda_j$ and given by the sum
\be\label{prop}
\sum_{m=1}^\infty \Bigl[\frac{1+e^{-2\lambda_i}}{e^{\lambda_i}+e^{-\lambda_i}}\frac{1+e^{-2\lambda_j}}{e^{\lambda_j}+e^{-\lambda_j}}\Bigr]^m=\frac{1}{e^{\lambda_i+\lambda_j}-1}
\ee
The regions in which three or more cycles are meeting become new vertices; the minimum order of a vertex is therefore three in the new diagrammatic technique. Since the maximum number of edges is then $6g-6+3s$, the number of combinatorial types of these graphs is finite for fixed $g$ and $s$.

We therefore come to the lemma.

\begin{lemma}\label{lm:diagrams}\cite{ACNP}
The genus-$g$ term of  $s$-backbone case is given by the following
(finite!) sum of homotopically equivalent diagrams (fatgraphs):
\beq
\Bigl\langle\prod_{i=1}^s\tr\log(x_i-H)\Bigr\rangle_g
=\sum_{{\mathrm{all\ genus\ } g}\atop \mathrm{graphs\ }\Gamma\mathrm{\ with\ }s\mathrm{\ faces}}
\frac{1}{\#\text{\rm Aut\,} (\Gamma)} \prod_{\mathrm{all\ edges}}\frac{1}{e^{\lambda_e^{(1)}+\lambda_e^{(2)}}-1}
:= F^{\text{Lag}}_g(\lambda_1,\dots,\lambda_s),
\label{new-model}
\eeq
where we allow all possible diagrams with vertices of order 3 and higher and we have exactly $s$ boundary components (faces). The factor $\#\hbox{Aut\,} (\Gamma)$ is the standard symmetry factor and the variables $\lambda_e^{(1)}$ and $\lambda_e^{(2)}$ are $\lambda$-variables of faces incident to the corresponding edge.
\end{lemma}

For $x_i=e^\lambda+e^{-\lambda}$, the point $\lambda=+\infty$ corresponds to the infinity point for $x$-variable on the physical sheet and  $\lambda=-\infty$ corresponds to the infinity point of $x$ on the unphysical sheet. Note also that $dx=0$ at $\lambda=0$ and at $\lambda=i\pi$, which are the corresponding two branching points.

Let us define the new times
\be
\label{TT}
T^\pm_k(\lambda):=\frac{1}{(2k+1)!!}\frac{1}N\sum_{i=1}^N\frac{\partial^{2k}}{\partial \lambda_i^{2k}}\frac{1}{\pm 1-e^{\lambda_i}},\quad k=0,1,\dots
\ee
A remarkable result of \cite{Ch95}, \cite{ABCO} is that $\forall g,s$ such that $2g-2+s>0$,
\be
\label{WT}
W^{(g)}_s(\lambda_\bullet)=\hbox{Pol\,}\bigl(\partial_{\lambda_\bullet}T^\pm_k(\lambda_\bullet)\bigr)d\lambda_1\cdots d\lambda_s,\quad k\le 2g+s-1,
\ee
so all $W^{(g)}_s$ are finite polynomials in the new times $T^\pm_k$. The same structure (\ref{WT}) holds \cite{ChN} for the Legendre and Laguerre ensembles. 

The times $T^\pm_k$ were identified with \emph{local times} in the topological recursion method. They are KdV hierarchy times for $\tau$-functions located at zeros of $dx$ \cite{DNoTop}, \cite{DNoTop1}, and they are also times appearing in the abstract topological recursion setting of \cite{ABCO}. Note a simple relation,
$$
\int_{+\infty}^{-\infty} \frac{\partial^{2k+1}}{\partial\lambda^{2k+1}}\frac{1}{\pm1-e^\lambda}d\lambda =\pm 1\cdot\delta_{k,0}.
$$

\begin{remark}
In cases of classic ensembles where $W^{(g)}_s$ admits representation (\ref{WT}), the virtual Euler characteristic is
$$
\varkappa_{g,s}=\hbox{Pol\,}\bigl(\partial_{\lambda_\bullet}T^\pm_k(\lambda_\bullet)\bigr)\Bigm|_{\partial_{\lambda_\bullet}T^\pm_k(\lambda_\bullet)=\pm 1\cdot\delta_{k,0}}.
$$
\end{remark}

\subsection{Relation to the virtual Euler characteristics}
A simple but crucial observation is that 
\be
\lim_{\lambda_e^{(1)},\lambda_e^{(2)}\to +\infty} \frac{1}{e^{\lambda_e^{(1)}+\lambda_e^{(2)}}-1}=0\quad \hbox{and}\quad 
\lim_{\lambda_e^{(1)},\lambda_e^{(2)}\to -\infty} \frac{1}{e^{\lambda_e^{(1)}+\lambda_e^{(2)}}-1}=-1,
\ee
and for any mixed limit when, say, $\lambda_e^{(1)}\to +\infty$ and $\lambda_e^{(2)}\to -\infty$, we assume that the first limit to $+\infty$ prevails, so the term $\frac{1}{e^{\lambda_e^{(1)}+\lambda_e^{(2)}}-1}$ again vanishes in this limit. We then observe from (\ref{kgs}) that 
\be
\varkappa_{g,s}=\frac{(-1)^s}{s!}\lim_{\lambda_i\to -\infty, \  i=1,\dots, s} F^{\text{Lag}}_g(\lambda_1,\dots,\lambda_s),
\ee
and since $F^{\text{Lag}}_g(\lambda_1,\dots,\lambda_s)$ vanishes if any of $\lambda_i$ tends to $+\infty$ and exploiting (\ref{loop-rep}), we obtain
\be\label{kgs-Ws}
\varkappa_{g,s}=\frac{(-1)^s}{s!}\underbrace{\int_{+\infty}^{-\infty}\cdots \int_{+\infty}^{-\infty}}_{s} W_s^{(g)}(\lambda_\bullet)
\ee
and finally, due to the Seiberg--Witten relations (\ref{SW-t0-mult}), we come to the following lemma.
\begin{lemma}\label{lm:kgs}
The virtual Euler characteristics for the Gaussian model is given by the following formula
\be
\varkappa^{\text{GUT}}_{g,s}=\left.\frac{(-1)^s}{s!} \frac{\partial^s F^{\text{Free-GUT}}_g}{\partial t_0^s}\right|_{t_0=1}
\ee
\end{lemma}

\subsection{Calculating $\varkappa_{g,s}$ for the Gaussian model}\label{ss:Gauss-kgs}
The asymptotic expansion for the Gaussian integral is well-known:
$$
\prod_{i=1}^{N}\Bigl[\int_{-\infty}^{\infty} dh_i\Bigr] \prod_{i<j}(h_i-h_j)^2e^{-\frac12 \sum_{i=1}^N h_i^2}=\prod_{k=0}^{N-1}k!
$$
and if we renormalize $N\to t_0 N$ with $N^{-2}$ a formal expansion parameter and $t_0$ a dynamical variable of the particle number, then we have to replace the product of factorials, or Gamma functions, by the Barnes function $G(z)$ \cite{Barnes} enjoying the difference equation
\be
G(z+1)=\Gamma(z)G(z)
\ee  
and having the asymptotic expansion \cite{Barnes} as $z\to\infty$
\be\label{G-as}
\log G(z+1)=z^2\Bigl[\frac12 \log z-\frac34\Bigr]+\frac12\log(2\pi)z-\frac{1}{12}\log z+\xi'(-1)+\sum_{k=1}^\infty \frac{B_{2k+2}}{4k(k+1)}z^{-2k},
\ee
with $B_{2k}$ being the Bernoulli numbers generated by the function
\be
\frac{1}{e^\lambda-1}=\frac1{\lambda}-\frac12+\sum_{k=1}^\infty \frac{B_{2k}}{(2k)!}\lambda^{2k-1},
\ee
with several first numbers $B_2=1/6$, $B_4=-1/30$, $B_6=1/42$, $B_8=-1/30$, $B_{10}=5/66$, etc.

We therefore identify 
\be\label{F-Gauss}
F^{\text{free-GUT}}(t_0N)=\log G(t_0N+1),
\ee
and considering only stable contributions $2g-2+s>0$, from the asymptotic expansion (\ref{G-as}) and Lemma~\ref{lm:kgs} we obtain the classic results that, for moduli spaces of oriented Riemann surfaces governed by the Gaussian Unitary Ensemble, 
\be
\label{Herm}
\varkappa^{\text{GUT}}_{0,s}=\frac{(s-3)!}{s!},\ s\ge 3;\qquad \varkappa^{\text{GUT}}_{g,s}=B_{2g}\frac{(2g+s-3)!}{s!(2g-2)!2g}(-1)^s,\ g\ge 1, \ s\ge 1.
\ee

\subsection{ $\varkappa^{\text{GUT}}_{g,s}$ via the Penner model}

It is well-known that the generating function for the virtual Euler characteristics $\varkappa^{\text{GUT}}_{g,s}$ of moduli spaces of genus-$g$ Riemann surfaces with $s$ punctures is given by the {\it Penner matrix model} \cite{Penner}
\be\label{Penner}
\sum_{g,s}\varkappa^{\text{GUT}}_{g,s}N^{2-2g}\hbar^{2-2g-s}=\log\int DH e^{N\hbar \tr[\log(1+H)-H]}-\log Z^{\text{free-GUT}},
\ee
where the integral is over Hermitian $N\times N$ matrices and is understood as an asymptotic expansion around the stable point $H=0$. Disregarding exponential correction not affecting the Laurent polynomial part of this expansion, we can rewrite the same integral as a generalized Laguerre ensemble integration:
\be\label{Lag-gen}
Z^{\text{gen-Lag}}(N,\hbar):=\int_{0}^{\infty} \prod_{i<j}(\lambda_i-\lambda_j)^2 \prod_{i=1}^{N} \lambda_i^{\hbar N}e^{-\sum_{i=1}^N \lambda_i}.
\ee  
From the generalized Laguerre polynomials $L_n(\lambda)=(-1)^n\lambda^{-N\hbar}e^{\lambda}\frac{d^n}{d^n\lambda}(\lambda^{N\hbar+n}e^{-\lambda})$ we obtain that $h_n=n! \Gamma(N\hbar +n+1)$ and again replacing $N$ by $t_0N$ and using the Barnes function, we obtain that
\be
Z^{\text{gen-Lag}}(N,\hbar)=G(N+1)(G(N+1+\hbar N)/G(\hbar N+1))
\ee
Taking now into account the normalization factor $\log Z^{\text{free-GUT}}=\log G(N+1)$ in (\ref{Penner}), we obtain that
\begin{align}
\sum_{g,s}\varkappa^{\text{GUT}}_{g,s}N^{2-2g}\hbar^{2-2g-s}&=\log G((1+\hbar)N+1)-\log G(\hbar N+1)\nonumber\\
&=\sum_{g=1}^\infty \sum_{s=1}^\infty \frac{\hbar^{2-2g-s}}{s!} \frac{\partial^s}{\partial t_0^s} F^{\text{free-GUT}}(t_0N)\mid_{t_0=1},
\end{align}
in full agreement with the statement of Lemma~\ref{lm:kgs}.

\section{Virtual Euler characteristics for other models}

\subsection{The Legendre model}\label{ss:Leg}
One of the main objectives of this text is to apply Lemma~\ref{lm:kgs} to the Legendre model defined as the model of orthogonal polynomials on the interval $[-2,2]$ with the unit measure: the corresponding polynomials are $\frac{n!}{(2n)!}\frac{\partial^n}{\partial \lambda^n} (\lambda^2-4)^n$,
$$
h_n=\frac{[n!]^4 2^{4n+2} }{(2n)!(2n+1)!},\qquad Z^{\text{Leg-free}}(N)=\prod_{n=0}^{N-1}h_n,
$$
so, for the partition function, we obtain
\be\label{Z-Leg}
Z^{\text{Leg-free}}(t_0N)=\frac{G^4(t_0N+1)2^{2(t_0N)^2}}{G(2t_0N+1)},
\ee
and using the asymptotic expansion for the Barnes function, we come to the following asymptotic expansion:
\be\label{F-Leg}
\log Z^{\text{Leg-free}}(t_0N)=Nt_0\log(2\pi)+3\xi'(-1)+\frac{1}{12}\log 2-\frac14\log(t_0N)+
\sum_{k=1}^\infty \frac{B_{2k+2}}{k(k+1)}\Bigl[1-\frac{1}{2^{2k+2}}\Bigr]\frac{1}{(t_0N)^{2k}} ,
\ee
and, in particular, for $\varkappa_{g,1}$ in the Legendre case, we have the following lemma.

\begin{lemma}\label{lm:kgs-Leg}
The virtual Euler characteristics $\varkappa^{\text{Leg}}_{g,1}$ for the Legendre ensemble reads
$$
\varkappa^{\text{Leg}}_{1,1}=-\frac14,\qquad \varkappa^{\text{Leg}}_{g,1}=\frac {-2B_{2g}}{g}\Bigl[1-\frac{1}{2^{2g}}\Bigr],\quad g\ge 2.
$$
\end{lemma}
This result coincides with that obtained by Norbury in \cite{Nor18} using the Toda chain hierarchy relations. In Sec.~\ref{s:one} we also derive several first terms $\varkappa^{\text{Leg}}_{g,1}$ using the Harer--Zagier type recursion for $W^{(g)}_1(\lambda)$.

For the Legendre model, we are not aware of any calculation of $\varkappa^{\text{Leg}}_{g,s}$ using Penner-like models.

\subsection{The model for nonorientable surfaces}

Our accounting for virtual Euler characteristics is not necessarily bounded by classical ensembles and classic orthogonal polynomials. A generalization to moduli spaces of nonorientable surfaces is provided by the Grand Orthogonal Ensemble (GOE), i.e., by integrals over symmetric real-valued matrices. 

The motivation of introducing the virtual Euler characteristics for nonorientable surfaces in \cite{ChZab} was based on combinatorial description of free-energy terms of $1/N$ expansion of matrix integrals over real-valued symmetric matrices. It was later discussed by  Goulden, Harer and Jackson \cite{GHJ} with the same motivation in mind. If we consider the integral over $N\times N$ symmetric real-valued matrices $R$ with the measure $DR=\prod_{i}dR_{i,i}\prod_{i<j}d R_{i,j}$, the correlation function 
$$
\langle R_{i,j} R_{k,l}\rangle :=\frac{1}{Z_0}\int R_{i,j} R_{k,l} e^{-\frac{N}{2}\tr R^2}DR=\delta_{j,k}\delta_{i,l}+\delta_{j,l}\delta_{i,k}\sim
\frac {1}{N}\raisebox{-12pt}{\mbox{
\begin{pspicture}(-0.7,-0.5)(0.7,0.5)
\pcline[linewidth=12pt,linestyle=solid](-0.5,0)(0.5,0)
\pcline[linewidth=10pt,linecolor=white,linestyle=solid](-0.52,0)(0.52,0)
\rput(-0.55,0.3){\makebox(0,0)[cr]{$i$}}
\rput(-0.55,-0.3){\makebox(0,0)[cr]{$j$}}
\rput(0.55,0.3){\makebox(0,0)[cl]{$l$}}
\rput(0.55,-0.3){\makebox(0,0)[cl]{$k$}}
%
\end{pspicture} 
}}
+
\frac {1}{N}\raisebox{-12pt}{\mbox{
\begin{pspicture}(-0.7,-0.5)(0.7,0.5)
\pcline[linewidth=1pt,linestyle=solid](-0.5,0.25)(0.5,-0.25)
\pcline[linewidth=4pt,linecolor=white,linestyle=solid](-0.52,-0.25)(0.52,0.25)
\pcline[linewidth=1pt,linecolor=black,linestyle=solid](-0.5,-0.25)(0.5,0.25)
\rput(-0.55,0.3){\makebox(0,0)[cr]{$i$}}
\rput(-0.55,-0.3){\makebox(0,0)[cr]{$j$}}
\rput(0.55,0.3){\makebox(0,0)[cl]{$l$}}
\rput(0.55,-0.3){\makebox(0,0)[cl]{$k$}}
%
\end{pspicture} 
}}
$$
can be depicted as an edge of a fat graph and the twisted edge with indices running along the sides; the vertices $\frac 1n \tr R^n$ then have the same form as in the Hermitian case. If we interpret index lines as faces of the simplicial decomposition of a new, nonoriented surface, then its Euler characteristics may take also odd integer values, but the corresponding matrix model has the same Penner-like form (\ref{Penner}) with the Hermitian matrix $H$ replaces by the real symmetric matrix $R$. 

The virtual Euler characteristics $\varkappa_{g,s}$ in the case of integration over real symmetric matrices is nonzero for integer and half-integer values of $g$ and $\varkappa^{\text{GOE}}_{g,s}=\varkappa^{\text{GUE}}_{g,s}+\varkappa^{\text{non-Or}}_{g,s}$, where $\varkappa^{\text{GUE}}_{g,s}$ are given by formula (\ref{Herm}) and  we have segregated the contribution due to the non-orientable surfaces.

\subsubsection{Nonorientable surfaces via Penner-like model}
The part $\varkappa^{\text{non-or}}_{g,s}$ was first calculated in \cite{ChZab} using the logarithmic model effectively governed by the eigenvalue model integral
\be\label{beta}
\int_{0}^\infty \prod_{i=1}^N d\lambda_i |\Delta(\lambda)|^{2\beta} \prod_{i=1}^N \Bigl[\lambda_i^{\hbar N} e^{-\hbar N\lambda_i}\Bigr],
\ee
with $\beta=1/2$, where $\Delta(\lambda)$ is the Vandermonde determinant. In  \cite{ChZab}, this integral was found by constructing a system of skew-orthogonal polynomials for the respective GOE (which were found to be closely related to the classic orthogonal polynomials of the generalized Laguerre type); the answer obtained was
\be\label{non-or}
\sum_{g\in \mathbb Z_+/2}\sum_{\hbar=1}^\infty \varkappa^{\text{non-Or}}_{g,s} N^{2-2g}\hbar^s
 =\frac{N}{2}(1+\hbar)\log(1+\hbar) -\sum_{k=1}^\infty \frac{2^{2k-1}-1}{2k(2k-1)}B_{2k}N^{1-2k}\Bigl[ \frac{1}{(\hbar+1)^{2k-1}}-1\Bigr].
\ee
This formula implies that only noninteger part of the virtual Euler characteristics (odd powers of $t_0$) gets corrections as compared with the Gaussian model. A similar formula was obtained by Goulden, Harer, and Jackson in \cite{GHJ} for arbitrary values of $\beta$ in the integral (\ref{beta}) using the Selberg integral technique. 

We now demonstrate that the corrections in (\ref{non-or}) are again in a perfect agreement with the statement of Lemma~\ref{lm:kgs}.

\subsubsection{Nonorientable surfaces via the Selberg integral}
We now consider the free $\beta$-model given by the Selberg integral
\be\label{Selberg}
Z_{\beta}:=\int_{-\infty}^\infty |\Delta(x)|^{2\beta} \prod_{j=1}^N e^{-x_j^2/2}dx_j=(2\pi)^{N/2}\prod_{j=1}^N \frac{\Gamma(1+\beta j)}{\Gamma(1+\beta)}.
\ee
We concentrate on the case $\beta=1/2$. In terms of the Barnes $G$-function, omitting irrelevant factors, the integral (\ref{Selberg}) then becomes
\be
Z^{\text{GOE-free}}=Z_{\beta=1}\simeq G(Nt_0+1)G(Nt_0+1/2).
\ee
In order to find a convenient asymptotic expansion, we use the product formula for the Barnes functions, $G(2z)\simeq G(z)G^2(z+1/2)G(z+1)$, which, upon substitution $z=t_0N+1/2$, gives
\be
G(2t_0N+1)=G(t_0N+1/2)G^2(t_0N+1)G(t_0N+3/2)= \bigl[ G(t_0N+1/2)G(t_0N+1)\bigr]^2 \Gamma (t_0N+1/2),
\ee
and we finally use the multiplication formula 
$$
\Gamma(t_0N)\Gamma(t_0N+1/2)=2^{1-2t_0N} \sqrt{\pi} \Gamma(2t_0N)
$$
to obtain
$$
\bigl[ G(t_0N+1/2)G(t_0N+1)\bigr]^2 = G(2t_0N+1)\Gamma(t_0N)/\Gamma(2t_0N).
$$
The partition function of the free $\beta=1$ model then reads:
\be
\log Z^{\text{GOE-free}}=\frac12 \log G(2t_0N+1) + \frac12 \log\Gamma (t_0N)-  \frac12 \log\Gamma (2t_0N).
\ee
We use the $z\to+\infty$ asymptotic expansion of the gamma function,
\be
\log \Gamma(z)=z\log z-z-\frac12 \log z+\sum_{k=1}^\infty \frac{B_{2k}}{2k(2k-1)}z^{1-2k},
\ee
and rescaling $2t_0N\to t_0N$ to obtain the relation
$$
2\log Z^{\text{free-GOE}}(t_0N)=\log Z^{\text{free-GUE}}(t_0N)-\log \Gamma(t_0N)+\log\Gamma(t_0N/2),
$$
and the correction due to non-orientable part (two last terms in the right-hand side) is
\be
F^{\text{non-Or}}(t_0N):= \sum_{k=0}^\infty \frac{B_{2k+2}}{(2k+2)(2k+1)}(2^{2k+1}-1)\frac{1}{(t_0N)^{2k+1}},
\ee
which produces the same expressions for $\varkappa^{\text{non-Or}}_{g,s}$ as formula (\ref{non-or}).

\subsection{The Marchenko--Pastur ensemble}
We now consider the model of rectangular $(1+\gamma)N\times N$ complex matrices $B$. It is well known that the free-field integral of this model can be presented in terms of (positive) eigenvalues $x_i$ of the Hermitian matrix $X=B^\dagger B$:
\be
Z^{\text{free-MP}}=\int DB\,DB^\dagger e^{-N\tr B^\dagger B}=
\int_0^{\infty} \Delta(x)^2 \prod_{i=1}^{N} \bigl[ x_i^{\gamma N}e^{-N x_i} dx_i \bigr]=\prod_{k=0}^{N-1} \Gamma(k+1)\Gamma(\gamma N+k+1).
\ee
Rescaling $N\to t_0N$ and doing the product, we obtain
\be
Z^{\text{free-MP}}=G(t_0N+1)G((1+\gamma)t_0N+1),
\ee
for instance, when $\gamma=0$, the integral is just the square of the GUT integral (which follows immediately from that $B$ is then a general square complex matrix, which can be written as a sum of two Hermitian matrices $Y$ and $Z$: $B=Y+iZ$ and $\tr B^\dagger B=\tr Y^2+\tr Z^2$, so we have a product of two $Z^{\text{free-GUT}}$). 

The situation becomes different if we consider the ``logarithmic'' integral:
\begin{align}
Z^{\text{log-MP}}&=\frac{1}{Z^{\text{free-MP}}}\int DB\,DB^\dagger e^{-\alpha N\tr (B^\dagger B)-\frac{\alpha N}{2} \tr (B^\dagger B)^2 -\frac{\alpha N}{3} \tr (B^\dagger B)^3 -\cdots}\nonumber\\
&=\frac{1}{Z^{\text{free-MP}}}\int_0^{``\infty''}  \Delta(x)^2 \prod_{i=1}^N \bigl[ dx_i x_i^{\gamma N}e^{\alpha N\log(1-x_i)}\bigr] \nonumber\\
&=\frac{1}{Z^{\text{free-MP}}}\int_0^1 \Delta(x)^2 \prod_{i=1}^N \bigl[ dx_i x_i^{\gamma N}(1-x_i)^{\alpha N}\bigr],
\end{align}
so, effectively, we obtain the integral of the generalized Legendre model! (Note that asymptotic expansion in $N^{2-2g}$ is not affected by changing the integration limits.) The monic orthogonal polynomials of this model are
$$
P_k(x)=x^{-\gamma N}(1-x)^{-\alpha N}\frac{d^k}{dx^k}\bigl[x^{\gamma N+k}(1-x)^{\alpha N+k}\bigr]\frac{\Gamma(\alpha N+\gamma N+k+1)}{\Gamma(\alpha N+\gamma N+2k+1)},
$$
and after performing a beta-function integration on the interval $[0,1]$, we obtain that 
\be
Z^{\text{log-MP}}=\prod_{k=0}^{N-1}\left[ \frac{1}{k! \Gamma(\gamma N+k+1)}k! \frac{\Gamma(\gamma N+k+1) \Gamma(\alpha N+k+1)\Gamma(\gamma N+\alpha N+k+1)}{\Gamma(\gamma N+\alpha N + 2k+1)\Gamma(\gamma N+\alpha N + 2k+2)} \right],
\ee
so, finally,
\be\label{kappa-MP-log}
Z^{\text{log-MP}}=\frac{G(\gamma N+\alpha N + N+1)G(\alpha N + N+1)}{G(\alpha N +1)G(\gamma N+\alpha N + 2N+1)},
\ee
and were we adopt this expression as a generating function for $\varkappa^{\text{MP}}_{g,s}$, the results will be substantially different from those for $Z^{\text{free-MP}}$. The genus $g\ge 2$ term of  $\log Z^{\text{log-MP}}$ reads
\be\label{log-MP}
\frac 1{N^{2g-2}}\frac{B_{2g}}{2g(2g-2)} \left[ \frac{1}{(\gamma+\alpha +1)^{2g-2}} + \frac{1}{(\alpha +1)^{2g-2}} - \frac{1}{\alpha^{2g-2}} - \frac{1}{(\gamma+\alpha +2)^{2g-2}}  \right],
\ee
whereas the same term in the free model is merely
\be\label{Chap}
\frac 1{(t_0N)^{2g-2}}\frac{B_{2g}}{2g(2g-2)}\left[ 1+ \frac{1}{(\gamma+1)^{2g-2}} \right]
\ee
lacking the fourth term in the brackets in (\ref{log-MP}). Note that, for $g=2$ and $\lambda=0$, formula (\ref{Chap}) gives $\varkappa_{2,0}=1/120$, which coincides with the result of Chapuy and Fang \cite{ChFang18}. 

\section{One-point function for the Legendre model}\label{s:one}

\subsection{Harer--Zagier-like linear recursion relations}\label{ss:HZ-Lag}

The derivation of the 1-point function is standard using orthogonal polynomials.
\allowdisplaybreaks
\begin{align*}
\left\langle {\rm tr} \left(\frac{1}{x-M}\right)\right\rangle&=
\int^1_{-1}\cdots \int^1_{-1}\sum_{i=1}^N\frac{1}{x-x_i}\Delta^2(x_1,...,x_N)dx_1\cdots dx_N\\
&=N\int^1_{-1}\cdots \int^1_{-1}\frac{1}{x-x_1}\Delta^2(x_1,...,x_N)dx_1\cdots dx_N\\
&=N\int^1_{-1}\frac{dx_1}{x-x_1}\prod_{k=2}^N\int^1_{-1}dx_k(x_1-x_k)^2\Delta^2(x_2,...,x_N)\\
&=N\int^1_{-1}\frac{dx_1}{x-x_1}\prod_{k=2}^N\int^1_{-1}dx_k\lim_{x'_1\to x_1}\frac{1}{x_1-x'_1}(x_1-x'_1)(x_1-x_k)(x'_1-x_k)\Delta^2(x_2,...,x_N)\\
&=N\int^1_{-1}\frac{dx_1}{x-x_1}\prod_{k=2}^N\int^1_{-1}dx_k\Delta(x_2,...,x_N)
\lim_{x'_1\to x_1}\frac{1}{x_1-x'_1}\Delta(x_1,x'_1,x_2,...,x_N)\\
&=N\int^1_{-1}\frac{dx_1}{x-x_1}\prod_{k=2}^N\int^1_{-1}dx_k\det\left[ p_{i-1}(x_j)\right]\lim_{x'_1\to x_1}\frac{1}{x_1-x'_1}\det\left[\begin{array}{c}p_{i-1}(x_1)\\p_{i-1}(x'_1)\\p_{i-1}(x_j)\end{array}\right]\\
&=N!\int^1_{-1}\frac{dx_1}{x-x_1}\det\left[\begin{array}{c}p_{i-1}(x_1)\\p_{i-1}(x'_1)\\ \int^1_{-1}p_{i-1}(x)p_{j-1}(x)dx\end{array}\right]\\
&=N!\prod_{i=0}^{N-3}r_i\int^1_{-1}\frac{dx_1}{x-x_1}\left[ p_{N-1}(x_1)p'_N(x_1)-p'_{N-1}(x_1)p_N(x_1)\right]
\end{align*}

\subsection{Five-term recursion}
Put $W_1(x)=\sum N^{2-2g}W_1^{(g)}(x)$.  So $W_1^{(g)}(x)=\omega_{g,1}(z)/x'(z)$.  This satisfies a differential equation (see Appendix A for the proof):
\begin{equation} \label{5-term}
\left\{\frac{1}{4}\frac{\partial^3}{\partial x^3}+\frac{2x}{x^2-4}\frac{\partial^2}{\partial x^2}+\frac{5x^2-4}{2(x^2-4)^2}\frac{\partial}{\partial x}-\frac{N^2-1}{x^2-4}\left(\frac{x}{x^2-4}+\frac{\partial}{\partial x}\right)\right\}W_1(x)=0.
\end{equation}
This is equivalent to a 5-term recursion between $\epsilon_g(k-2)$, $\epsilon_g(k-1)$, $\epsilon_g(k)$, $\epsilon_{g-1}(k-1)$,  $\epsilon_{g-1}(k)$ where
$$ W_1^{(g)}=\sum\epsilon_g(k)x^{-2k-1}.$$
The differential equation (\ref{5-term}) on the one-point correlation function was obtained by Gaberdiel, Klemm, and Runkel \cite{GKR05} from conformal field theory considerations and by Norbury \cite{NoGW} from cohomological field theory standpoint.

It is convenient to rescale 
\be
\epsilon_g(k)=\frac{(2k)!}{k!\,k!} f_g(k).
\ee
Then, for $f_g(k)$, we have the five-term recursion relation
\be
-4k^2 f_g(k)+(8k^2-12k+6)f_g(k-1)-4(k-1)^2 f_g(k-2)+4(f_{g+1}(k)-f_{g+1}(k-1))=0.
\ee
In particular, $f_0(k)=1$ for all $k\ge 0$, so we have the standard result $W_1^{(0)}=\frac1{\sqrt{x^2-4}}dx=\sum_{k=0}^\infty \frac{(2k)!}{k!\,k!}\frac {dx}{x^{-2k-1}}$.

\subsection{$u$-variables}

It is very convenient to express correlation functions using different variables: For $g\ge1$, the 1-point correlation function 
$$
W_1^{(g)}(x)=\sum_{n=0}^{g-1}r_n^{(g)}u_n(\lambda)dx=\sum_{n=0}^{g-1}r_n^{(g)}u_n(\lambda)\cdot(e^\lambda-e^{-\lambda})d\lambda,
$$ 
where
$$
u_n(\lambda):=\frac1{(e^\lambda-e^{-\lambda})^{3+2n}},\qquad x=e^\lambda+e^{-\lambda}.
$$
The recursion equation (\ref{5-term}) takes an especially simple form in terms of the coefficients $r_n^{(g)}$:
\begin{equation} \label{b-recursion}
(2n+2)r_n^{(g)}=(2n+1)^2\Bigl[\frac14 (2n+2) r_n^{(g-1)}+(2n-1)r_{n-1}^{(g-1)}\Bigr]
\end{equation}
with $r_0^{(1)}=1$. This recursion is very close to the one in \cite{ACNP} for the Gaussian means; in particular, for the boundary terms it is just a two-term recursion which immediately yields
$$
r_0^{(g)}=4^{-g+1}r_0^{(1)},\qquad r_{g-1}^{(g)}=\frac{\bigl[(2g-1)!!\bigr]^2(2g-3)!!}{(2g)!!}r_0^{(1)}.
$$
If we rescale $r_n^{(g)}=4^{-g+1}\tilde r_n^{(g)}$, then, presumably, \emph{all $\tilde r_n^{(g)}$ are positive integers for $0\ge n\ge g-1$}.

Using the identity
$$
\frac{4k(k+1)}{(e^\lambda-e^{-\lambda})^{k+2}}=\Bigl[\frac{d^2}{d\lambda^2}-k^2 \Bigr]\frac{1}{(e^\lambda-e^{-\lambda})^{k}}
$$
we immediately obtain that 
$$
\int_{-\infty}^{\infty} \frac{1}{(e^\lambda-e^{-\lambda})^{2k+2}}d\lambda=\frac{(-1)^{k+1}}{2}\cdot \frac{(k!)^2}{(2k+1)!},
$$
where we used that 
$$
\int_{-\infty}^{\infty} \frac{1}{(e^\lambda-e^{-\lambda})^{2}}d\lambda=-\frac{1}{4}\int_{-\infty}^{\infty} \frac{d}{d\lambda}\frac{e^\lambda+e^{-\lambda}}{e^\lambda-e^{-\lambda}}d\lambda=-\frac12
$$
independently on the choice of a path circumnavigating the pole at $\lambda=0$.

\subsection{Comparing with $\varkappa^{\text{Leg}}_{g,1}$}

We now evaluate $W_1^{(g)}(\lambda)$ for the first few $g$ to compare with the results of Lemma~\ref{lm:kgs-Leg}. Denoting
\be
\kappa_{n}^{(g)}:=r_n^{(g)}\int_{-\infty}^{\infty} \frac{1}{(e^\lambda-e^{-\lambda})^{2n+2}}d\lambda=r_n^{(g)}\frac{(-1)^{n+1}}{2}\cdot \frac{(n!)^2}{(2n+1)!},
\ee
we obtain the three-term recursion relation for $\kappa_{n}^{(g)}$:
\be
4(n+1)\kappa_{n}^{(g)}=(2n+1)^2(n+1)\kappa_{n}^{(g-1)}-(2n+1)n(2n-1)\kappa_{n-1}^{(g-1)},
\ee

If we begin with $\kappa_0^{(1)}=-1/2=2\varkappa^{\text{Leg}}_{1,1}$, then for genus $2$ we have $2\varkappa^{\text{Leg}}_{2,1}=\kappa_0^{(2)}+\kappa_1^{(2)}=-\frac18+\frac{3}{16}=\frac{1}{16}$, for genus $3$ we have $2\varkappa^{\text{Leg}}_{3,1}=\kappa_0^{(3)}+\kappa_1^{(3)}+\kappa_2^{(3)}=-\frac1{32}+\frac{30}{64}-\frac{15}{32}=-\frac{1}{32}$, and for genus $4$ we have $2\varkappa^{\text{Leg}}_{4,1}=\kappa_0^{(4)}+\kappa_1^{(4)}+\kappa_2^{(4)}+\kappa_3^{(4)}=-\frac1{128}+\frac{3\cdot 7\cdot 13}{2^8}-\frac{3\cdot 7\cdot 25}{2^7}+\frac{7\cdot 9\cdot 25}{2^9}=\frac{17}{2^9}$ in full agreement with Lemma~\ref{lm:kgs-Leg}.


\setcounter{section}{0}
\appendix{Derivation of (\ref{5-term})}

For $W_1(x)$ we have a general formula in terms of orthogonal polynomials:
$$
W_1(x)=\int_{-2}^2 dx_1 \frac{L_N'(x_1)L_{N-1}(x_1)-L_{N-1}'(x_1)L_N(x_1)}{x-x_1},
$$
where $L_N(x_1)$ are the Legendre polynomials satisfying the standard differential equation
\beq
(x_1^2-4)\frac{d^2}{dx_1^2} L_N(x_1)-2x_1 \frac{d}{dx_1} L_N(x_1)+N(N+1) L_N(x_1)=0.
\label{diff}
\eeq
We introduce the quantity (and suppress in what follows the argument of expressions wherever possible)
\beq
Q(x):=(x^2-4)(L'_N(x)L_{N-1}(x)-L'_{N-1}(x)L_N(x))
\eeq
We now find the differential equation on $Q$ using (\ref{diff}):
\begin{align}
&Q'=((x^2-4)L'_N)'L_{N-1}-((x^2-4)L'_{N-1})'L_N = 2N L_N L_{N-1}\nonumber\\
&Q''=2N(L'_NL_{N-1}+L_NL'_{N-1})\nonumber\\
&Q'''=4N L'_NL'_{N-1} +\frac{2N}{x^2-4}\bigl[ -2x(L'_NL_{N-1}+L_NL'_{N-1}) +2N^2 L_NL_{N-1} \bigr]=4N L'_NL'_{N-1} -\frac{2x}{x^2-4}Q'' +\frac{2N^2}{x^2-4}Q'\nonumber\\
&4N(L'_NL'_{N-1})'=\frac{4N}{x^2-4}\bigl[ -4xL'_NL'_{N-1} +N^2 (L'_NL_{N-1} +L'_{N-1}L'_N)-N (L'_N L_{N-1}-L'_{N-1}L_N)\bigr]\nonumber\\
&\qquad = -\frac{4x}{x^2-4} \Bigl[ Q''' +\frac{2x}{x^2-4} Q'' -\frac{2N^2}{x^2-4}Q' \Bigr] + \frac{2N^2}{x^2-4}Q'' -\frac{4N^2}{(x^2-4)^2}Q,
\nonumber
\end{align}
so, finally, differentiating $Q'''$ once more, we obtain
\beq
Q^{\text{IV}}=-\frac{6x}{x^2-4}Q''' +\frac{-6x^2+8}{(x^2-4)^2}Q''+4N^2\Bigl[\frac{1}{x^2-4}Q'' +\frac{x}{(x^2-4)^2}Q'-\frac{1}{(x^2-4)^2}Q \Bigr].
\label{Q-4}
\eeq
In order to obtain a differential equation on $W_1(x)$ itself, we can integrate (\ref{Q-4}) for the variable $x_1$ with the weight $\frac{(x_1^2-4)^3}{x-x_1}dx_1$ from $-2$ to $2$ expressing derivatives integrating by parts. The weight is such that all boundary terms vanish, so we obtain
\begin{align*}
&\int_{-2}^2dx_1 Q(x_1)\Bigl[ \frac{\partial^4}{\partial x_1^4}\frac{(x_1^2-4)^3}{x-x_1} - \frac{\partial^3}{\partial x_1^3}\frac{6x_1(x_1^2-4)^2}{x-x_1} + \frac{\partial^2}{\partial x_1^2}\frac{(6x_1-8)(x_1^2-4)}{x-x_1}\Bigr]\\
&\qquad -4N^2 \int_{-2}^2dx_1 Q(x_1)\Bigl[ \frac{\partial^2}{\partial x_1^2}\frac{(x_1^2-4)^2}{x-x_1} - \frac{\partial}{\partial x_1}\frac{x_1(x_1^2-4)}{x-x_1} - \frac{x_1^2-4}{x-x_1}\Bigr]=0.
\end{align*}
Several simplifications follow from that $Q(x_1)$ is symmetric, so all integrals $\int_{-2}^2 dx_1 Q(x_1) x_1^{2k+1}$ vanish. We can check explicitly that all terms with even nonnegative powers of $x_1$ as $x_1\to\infty$ vanish. And we simplify the first line differentiating once the first summand and canceling with the second one thus obtaining
$$
\int_{-2}^2dx_1 Q(x_1)\Bigl[ \frac{\partial^3}{\partial x_1^3}\frac{(x_1^2-4)^3}{(x-x_1)^2} + \frac{\partial^2}{\partial x_1^2}\frac{(6x_1-8)(x_1^2-4)}{x-x_1}\Bigr]
$$
Upon transforming the first line and segregating the factor $(x_1^2-4)$ in $Q(x_1)$, we obtain
$$
\int_{-2}^2dx_1 (L_N'L_{N-1}-L_NL'_{N-1})\Bigl[ -\frac{\partial}{\partial x}\Bigl( 6(x_1^2-4)\frac{(x^2-4)^3}{(x-x_1)^4}\Bigr) +
2(x_1^2-4)\frac{(6x^2-8)(x^2-4)}{(x-x_1)^3} \Bigr],
$$
which can be already expanded into derivatives of $W_1(x)$. For the second line we have a similar transformation bringing it to
$$
-4N^2\int_{-2}^2dx_1 (L_N'L_{N-1}-L_NL'_{N-1})\Bigl[ 2\frac{(x^2-4)^3}{(x-x_1)^3} - \frac{5x(x^2-4)^2}{(x-x_1)^2}
+\frac{(3x^2-4)(x^2-4)}{x-x_1} \Bigr],
$$
so, after some algebra, we obtain a differential equation on $W_1(x)$:
\begin{align*}
&(x^2-4)^4 W^{\text{IV}}_1(x) +14x(x^2-4)^3 W'''_1(x) +(50x^2-40)(x^2-4)^2W''_1(x)+x(40x^2-96)(x^2-4)W'_1(x)\\
&-4(N^2-1)\Bigl[ (x^2-4)^3 W''_1(x) +5x(x^2-4)^2 W'_1(x) + (3x^2-4)(x^2-4)W_1(x)\Bigr]=0.
\end{align*}
We can simplify it remembering that in the planar limit we have $W^{(0)}_1(x)=(x^2-4)^{-1/2}$ and that it satisfies a linear differential equation
$$
\Bigl[\frac{d}{dx}+\frac{x}{x^2-4}\Bigr] W_1^{(0)}(x)=0.
$$
We then observe that the term in the second line can be transformed into
$$
-4(N^2-1)(x^2-4)^3\Bigl[\frac{d}{dx}+\frac{4x}{x^2-4}\Bigr]\Bigl[\frac{d}{dx}+\frac{x}{x^2-4}\Bigr]W_1(x)
$$
whereas the first line admits a similar factorization:
$$
(x^2-4)^3\Bigl[\frac{d}{dx}+\frac{4x}{x^2-4}\Bigr]\Bigl[(x^2-4)W'''_1(x) +8x W''_1(x) +\frac {10x^2-8}{x^2-4}W'_1(x) \Bigr].
$$
Because the kernel of the operator $\Bigl[\frac{d}{dx}+\frac{x}{x^2-4}\Bigr]$ are functions $(x^2-4)^{-2}$, which never appear in perturbative solutions of $W_1(x)$ (in which expansions are only in half-integer powers of $(x^2-4)$, we can eliminate this linear differential operator from the both lines thus claiming that the proper one-point correlation function must satisfy a simpler differential equation
$$
W'''_1(x)+\frac{8x}{x^2-4}W''_1(x)+\frac{10x^2-8}{(x^2-4)^2}W'_1(x)-\frac{4(N^2-1)}{x^2-4}\Bigl[W'_1(x)+\frac{x}{x^2-4}W_1(x)\Bigr]=0,
$$
that is, exactly (\ref{5-term}).

\end{document}